\documentclass[preprint,aps]{revtex4}
\usepackage{CJK}
\usepackage{amsmath,chemarrow}
\usepackage[dvips]{graphicx}
\usepackage{dcolumn}
\usepackage{bm}
\usepackage[tight,footnotesize]{subfigure}

\usepackage{color}
     \definecolor{darkred}{rgb}{0.75,0,0}
     \definecolor{darkgreen}{rgb}{0,0.5,0}
     \definecolor{darkblue}{rgb}{0,0,0.75}
     \definecolor{darkorange}{rgb}{1,0.9,0.1}

\begin{document}
\begin{CJK*}{GBK}{}




\title{Fixation, transient landscape and diffusion's dilemma in stochastic evolutionary game dynamics}

\author{Da Zhou $^{1, 2}$}
\email{zhouda1112@math.pku.edu.cn}
 \address{$^1$School of Mathematical Sciences, Peking University,
Beijing 100871, China}
 \address{$^2$MOE Key Laboratory of Bioinformatics and Bioinformatics Division, TNLIST/
 Department of Automation, Tsinghua University, Beijing 100864, China}

\author{Hong Qian $^3$}
\email{hqian@u.washington.edu}
 \address{$^3$Department of Applied Mathematics, University of Washington
Seattle, WA 98195, USA}

\begin{abstract}

Agent-based stochastic models for finite populations have recently received much attention in the game theory of evolutionary dynamics.
Both the ultimate fixation and the pre-fixation transient behavior are important to a full understanding of the dynamics.
In this paper, we study the transient dynamics of the well-mixed Moran process through constructing a {\em landscape} function.  It is shown that the
landscape playing a central theoretical ``device'' that integrates several lines of inquiries: the stable behavior
of the replicator dynamics, the long-time fixation, and continuous diffusion approximation associated with asymptotically large population. Several issues relating to the transient dynamics are discussed: 1) Multiple time scales phenomenon associated
with intra- and inter-attractoral dynamics; 2) discontinuous
transition in stochastically stationary process akin to
Maxwell construction in equilibrium statistical physics; and
3) the dilemma diffusion approximation facing as a continuous
approximation of the discrete evolutionary
dynamics. It is found that \emph{rare events} with exponentially small probabilities, corresponding to the uphill movements and barrier crossing in the landscape with multiple wells that are made possible by strong nonlinear dynamics, plays an important role in understanding the origin of the complexity in evolutionary, nonlinear biological systems.

\end{abstract}

\maketitle
\end{CJK*}

\section{Introduction}
\label{}

One of the salient features of the stochastic evolutionary game dynamics for finite populations is the \emph{fixation} \cite{Nowak book, Traulsen's review, Roca09PLR}. That is, no matter how the initial strategies are distributed in a population, the system will eventually be fixated to only one strategy. In general, this phenomenon can be theoretically explained in terms of a Markov process with absorbing state(s) \cite{Karlin book, Durrett book}: The limiting theory of Markov processes tells us that the finite-state Markov chains with absorbing states will eventually be trapped into one of the absorbing states as time passes. The well-known Wright-Fisher model \cite{Fisher1922on, Wright1931evolution}, Moran process \cite{Moran book, Nowak Nature 04} and the pairwise comparison processes \cite{Traulsen 07 Fermi, Fu10JTB} all belong to this class.  A further classification of Darwinian selection scenarios based on the fixation descriptions has already been established in stochastic game dynamics \cite{Taylor BMB 04, Antal BMB 06, Taylor06JTB, Ohtsuki JTB 07, Altock NTP 09, Wubin2010PRE}.
Another important class consists of evolutionary dynamics with \emph{mutations} where an ergodic mutation-selection equilibrium can be reached \cite{Fudenberg TPB 06, Antal09JTBa, Antal09JTBb}.  The latter
is out of the scope of our study; However, the {\em landscape} introduced
in the present work provides a unified perspective for both classes of processes.

In addition to the ultimate fixation, attention should also be paid to the time-dependent, pre-fixation transient behavior for several reasons
\cite{szabo2007evolutionary, DrozEPJ, SzaboPRE}:
On one hand, the process to fixation is intimately dependent upon both the transient movements before absorption and the one last step to fixation. That is, studying the transient dynamics provides important insights into the final fixation behavior. On the other hand, sometimes the time for a true fixation is too long to be observed \cite{Antal BMB 06} and the relevant time scale can be shorter \cite{Huisman Nature 99, Huisman Ecology 01, Dorroch 1965}. In this case, the transient dynamics provides a more appropriate description. Furthermore, examinations of transients can yield a mechanistic understanding of the \emph{persistence} and \emph{coexistence} in complex biological dynamics, especially in ecosystems \cite{Hastings 04}. Indeed, it has been found that the pre-fixation transient dynamics could be an essential explanatory aspect of characterizing the stochastic fluctuations raised from finite populations \cite{Block BMB 2000, Claussen 05 PRE, Tao BMB 07, Vellela BMB 07, Ficici JTB 07}.

The theory of quasi-stationarity is a widely applied, standard technique of studying the pre-fixation process \cite{quasibook}. It defines the subchain with the absorbing states removed. Based on this approach, we present an extended analysis for the transient dynamics of the well-mixed frequency-dependent Moran process. An ergodic \emph{conditional} stationary distribution is used to characterize the pre-fixation process.
As a result of the law of large numbers, this stationary distribution approaches to a singular distribution in the infinite population limit. The corresponding large deviation rate function \cite{Freidlin, Dembo, Touchette 10 PR}, which is population-size independent, is shown to be a landscape.
This \emph{transient landscape} has a Lyapunov property with respect to the corresponding deterministic replicator dynamics, providing a potential-like function for visualizing the transient stochastic dynamics. Ideas related to the transient landscape of Moran
process have been discussed in the past: Claussen and Traulsen \cite{Claussen 05 PRE} studied
non-Gaussian stochastic fluctuations based on the conditional stationary distribution. It is also a general feeling that one can use the
negative logarithm of the stationary, or conditional
stationary distribution as the potential in evolutionary
dynamics, following an analogue to Boltzmann's law in statistical
mechanics.
However, it is important to point out that a stationary
distribution usually collapses to singular supports in the
infinite population limit, while our large deviation rate function
$\psi(x)$ is supported on the whole space and it is independent
of system's size.  Therefore, in terms of the analogue to
Boltzmann's law, we are effectively identifying the system's
size as the inversed temperature which tends to infinity
for a deterministic limit.

Even though our analysis is based on the one-dimensional Moran
process, this idea is general. It can be applied to many other
multi-dimensional evolutionary game dynamics with finite populations,
with or without detailed balance \cite{qian_2011_nonlinearity_review}.
For the latter case, the landscape itself is an {\em emergent property} of the dynamics.  With respect to Moran process, \cite{Roca09PLR, Antal BMB 06} also discovered the expression of $\psi(x)$ from a different origin, via their approximated calculation of fixation probability for large population size. We shall show that this connection is a nice mathematical property of the
$\psi(x)$ function for the processes in one-dimensional case, but its
generalization to multi-dimensional cases is not obvious. More specifically,
for multi-dimensional systems with multiple
alleles, the fixation probability does not naturally give a landscape. The large deviation rate function, however, can be generalized to multi-dimensional
Markov processes, as indicated by the Freidlin-Wentzell theory \cite{Freidlin}.
The landscape we introduced is also consistent with the
landscape theory for other population dynamics, e.g., chemical,
that is ergodic without fixation \cite{Wang08PNAS, qian_2011_nonlinearity_review}.

There are two fundamentally different types of movements in this landscape that require separated attention.  ($i$) ``Downhill movements'' which have deterministic counterparts: The local minima (\emph{transient attractors}) in this landscape correspond to the stable points in the replicator dynamics \cite{Taylor and Jonker1978}. That is, these transient attractors are in direct agreement with the \emph{evolutionarily stable strategies} (ESSs) \cite{Maynard book, Zhou Da JTB 10}.  ($ii$) ``Uphill movements'' which are rare and without a deterministic correspondence. In general, rare events take exponentially long time; one needs to take multiple time scales into consideration in understanding the appropriate fluctuation descriptions for the transient dynamics as well as eventual fixation. This is particularly relevant in the anti-coordination games.

Furthermore, the concept of \emph{stochastic bistability} is studied in the coordination games.  In this case, the downhill and uphill movements in the landscape dominate ``intra-attractoral'' and ``inter-attractoral'' dynamics respectively \cite{qian_2011_nonlinearity_review}.
It is shown that a Maxwell-type construction from classic phase transition theory in statistical physics \cite{ge hao PRL 09} is necessary as the population size tends to infinity, i.e., only one of attractors should be singled out in such a construction | It corresponds to the global minimum of the landscape. This is not present in the bistable deterministic dynamics; it raises the novel issue of {\em ultimate fixation}.  It did not escape our notice that it is the {\em exponentially long-time search} that ultimately finds the global minimum in a ``non-convex optimization'' \cite{non convex}.

Another important issue directly relating to the transient dynamics is the diffusion approximation \cite{Feller1954diffusion, Gardiner book, tan2000stochastic}. With the conventional truncation of Kramers-Moyal (KM) expansion, the discrete stochastic Moran process for large populations has been approximated by a stochastic differential equation \cite{Traulsen PRL05, Chalub 08}, with absorbing Dirichlet boundary conditions. If one replaces the absorbing boundary conditions with the reflecting ones, we can also derive an ergodic stationary distribution from the Fokker-Planck equation of this diffusion process. It will be shown that this stationary distribution is in fact the ``conditional'' stationary distribution for the process with absorbing boundary conditions. In a comparison of the transient dynamics between the original Moran process and its continuous counterpart, it is shown that even though the KM diffusion is valid in finite time as a local dynamical approximation, it could lead to incorrect approximation in global inter-attractoral dynamics. In bistable game systems, particularly, the KM diffusion could single out a different stable point from that of the original process for large but finite populations. Moreover, enlightened by H\"{a}nggi et al.'s work \cite{hanggi1984bistable}, we also consider their diffusion approximation that provides the correct global dynamics. However, this diffusion process gives
incorrect finite time stochastic dynamics.

Now we have a \emph{diffusion's dilemma}: The truncated KM diffusion gives the correct finite time stochastic dynamics as the original Moran process with large population size (this is guaranteed both by the so called van Kampen's system size expansion \cite{van Kampen} and Kurtz's theorem \cite{Kurtz1976, Kurtz1978}), but wrong stationary distribution.  On the other hand,  H\"{a}nggi et al.'s diffusion, which is unique in providing the correct stationary distribution as well as deterministic limit, is wrong for the finite time stochastic dynamics. To further illustrate this diffusion's dilemma, a simple example is present. By investigating the first passage times,
it is found that the failure of exponential approximation in the uphill movement could be the origin of the difficulties of
diffusion approximation. In other words, diffusion approximation
is a second-order polynomial expansion for the Kolmogorov forward equation of the original discrete
process, which can give the correct Gaussian dynamics near the stable point; However, the inter-attractoral global dynamics, determined by the barrier crossing events with exponential small probabilities, should be approximated in the level of exponential asymptotics.

This paper is organized as follows: In Sec. II, we introduce the frequency-dependent Moran process.
Then we give the transient description of the Moran process in Sec. III, where the transient landscape $\psi(x)$ is
constructed. It is shown that this landscape as a ``glue'' holds the deterministic replicator dynamics, the fixation
and the problem of Maxwell-type construction together. Diffusion's dilemma is discussed in Sec. IV. The discussions
are included in the last section.

\section{Frequency-dependent Moran process}

To study evolutionary game theory in finite populations, Nowak et al. \cite{Nowak Nature 04} generalized Moran's classical population genetic model \cite{Moran book} by using frequency-dependent fitness. Consider a population of $N$ individuals playing a symmetric $2\times 2$ game with strategies $A$ and $B$, the payoff matrix is
\begin{equation}M=\left(\begin{array}{cc}a & b\\ c & d
\end{array}\right),\end{equation}
where all the entries in the matrix are assumed to be non-negative. If $i$ players follow strategy $A$, and $N-i$ play $B$, the the average payoff of an individual of $A$ is
\begin{equation}
F_A^i=\frac{a(i-1)+b(N-i)}{N-1},
\end{equation}
where self-interaction is excluded, and also for $B$ is
\begin{equation}
F_B^i=\frac{ci+d(N-i-1)}{N-1}.
\end{equation}
Fitness is assumed to be a linear combination of background fitness and the payoff as follows:
\begin{equation}
f_A^i=1-w+wF_A^i,
\end{equation}
\begin{equation}
f_B^i=1-w+wF_B^i,
\end{equation}
where $w\in[0,1]$ denotes the intensity of selection.

At each time step, one individual is chosen to reproduce proportional to its fitness, then the offspring replaces a randomly chosen individual, so that the population size is constant. The evolutionary mechanism in this Moran process, thus, can be conveniently described as a Markov chain. If we denote $X$ as the the number of individuals following strategy $A$, then $X$ is actually a finite-state birth-death process with discrete-time
steps, which can be expressed as follows:
$$0\overset{T_1^-}{\longleftarrow}1\underset{T_2^-}{\overset{T_1^+}{\rightleftharpoonsfill{20pt}}}2\rightleftharpoonsfill{20pt}\cdot\cdot\cdot
\rightleftharpoonsfill{20pt}(N-2)\underset{T_{N-1}^-}{\overset{T_{N-2}^+}{\rightleftharpoonsfill{20pt}}}(N-1)\overset{T_{N-1}^+}{\longrightarrow}N,$$
where the birth and death probabilities conditional on the present state $X=i$ are
\begin{equation}
T_i^+=P(i\rightarrow
i+1)=\frac{if_A^i}{if_A^i+(N-i)f_B^i}\frac{N-i}{N},
\end{equation}
\begin{equation}
T_i^-=P(i\rightarrow
i-1)=\frac{(N-i)f_B^i}{if_A^i+(N-i)f_B^i}\frac{i}{N},
\end{equation}
\begin{equation}
T_i^0=P(i\rightarrow i)=1-T_i^+-T_i^-.
\end{equation}
So the Markov transition probability matrix of this process can be denoted as
\begin{equation}
Q=\left(\begin{array}{ccccc}
1&0&0&\cdots&0\\
T_1^- & 1-(T_1^++T_1^-) & T_1^+ & \cdots & 0 \\
\vdots & \vdots & \vdots & \vdots & \vdots  \\
0 & \cdots & T_{N-1}^- & 1-(T_{N-1}^++T_{N-1}^-) & T_{N-1}^+ \\
0 & \cdots & 0 & 0 & 1 \\
\end{array}\right).
\end{equation}
It is easy to see that $X=0$ and $N$ are two absorbing boundaries for the system, so the limiting distribution of the Moran process can be denoted as
\begin{equation}
(1-\rho_i)\delta_0+\rho_i\delta_N,
\end{equation}
where $\rho_i$ is the fixation probability that $i$ individuals of strategy $A$ succeed in taking over the entire population, and $\delta_0$ (or $\delta_N$) is a Dirac mass at $0$ (or $N$). In other words, except for the two pure-strategic absorbing states, all the mixed states are transient. This process is different from an irreducible Markov chain whose ergodic behavior is supported on the whole state space. Thus in the Moran process, it is natural to investigate the conditions for selection to favor successful fixation by calculating the corresponding fixation probabilities \cite{Taylor BMB 04}.

\section{transient analysis}

The fixation is intimately dependent upon the transient behavior before absorption. To show this, we shall divide the process to fixation into two parts: One is the irreducible transitions among the transient states before absorption, the other is the last step to fixation.
This distinction has a very clear evolutionary meaning: the former is a consequence of ``global dynamics'' while the latter is a rather local event.  It is also worth mentioning that these two scenarios correspond so-called ``golf-course search'' and
``funnel perspective'' in the field of protein folding \cite{Qian02PS}.
By explicitly studying these two different problems, one is able to address the ``dynamic nature of fixation'': Is the dynamics toward fixation inherent in the evolutionary dynamics? The investigation of the transients provides another perspective to understand the fixation. Further, in order to describe the inherent fluctuations arising from finite populations, the transient dynamics has been proved to be an essential explanatory aspect \cite{Claussen 05 PRE, Tao BMB 07}.

\subsection{Conditional stationary distribution and the transient landscape}

A natural mathematical idea to describe the transient behavior is to concentrate on
the distribution conditional on the subspace of the mix-strategic states \cite{quasibook}.
We consider an auxiliary process $X^*(t)$ close to the original Moran process
$X(t)$. $X^*(t)$ can be described as the original process with the absorbing states
removed, while all other birth and death probabilities remain unchanged \cite{quasibook, Nasell 99 MB}.
So the process of $X^*(t)$ can be expressed as
$$1\underset{T_2^-}{\overset{T_1^+}{\rightleftharpoonsfill{20pt}}}2\rightleftharpoonsfill{20pt}\cdot\cdot\cdot
\rightleftharpoonsfill{20pt}(N-2)\underset{T_{N-1}^-}{\overset{T_{N-2}^+}{\rightleftharpoonsfill{20pt}}}(N-1).$$
Note that $X^*(t)$ is an ergodic Markov chain
with a unique stationary distribution $\{\eta_i\}$, yielding the following iterative relation
\begin{equation}
\eta_i=\eta_1\prod_{k=2}^i
		\left(\frac{T_{k-1}^+}{T_k^-}\right),
           ~~~~i=2,...,N-1.
\label{CSD}
\end{equation}

The interpretation of $\eta$ for the transient description of the original process $X(t)$
is from the theory of quasi-stationarity \cite{Dorroch 1965}.
It is shown that $\eta$ characterizes
the occupation time distribution of the process before absorption,
i.e. $\eta_i$ is the average times of visiting to $i$ divided by the mean absorbing time (see appendix A).
So we call $\eta$ the \emph{conditional stationary distribution} of the original process.

We write $\eta$ as $\eta^{(N)}$ for its dependence on the population size $N$,
$$\eta^{(N)}_i=\eta^{(N)}_1\prod_{k=1}^i\left(
\frac{T_{k-1}^+}{T_k^-}\right) =A\exp\left\{\sum_{k=1}^i\ln\left(\frac{T_{k-1}^+}{T_k^-}\right)\right\}.$$
By expanding in the inverse of large population size $N$, $\eta^{(N)}$ can be expressed as
$$\eta^{(N)}(x)=A\exp\left[-N\psi(x)+\psi_1(x)+\frac{\psi_2(x)}{N}+...\right]. ~~~x=\frac{i}{N}.$$
where the leading item $\psi(x)$ is the large deviation rate function of $\eta^{(N)}$
\cite{Dembo, Touchette 10 PR, Ge 10 PRSI}.
By the Euler-MacLaurin summation formula, we have
\begin{equation}
\psi(x)=-\int_0^{x}\ln\left[\frac{T^+(y)}{T^-(y)}\right]dy,
\label{M-potential}
\end{equation}
where
\begin{equation}
T^+(y)=\frac{x(1-x)(1-w+w(ax+b(1-x)))}{x(1-w+w(ax+b(1-x)))+(1-x)(1-w+w(cx+d(1-x)))},
\end{equation}
\begin{equation}
T^-(y)=\frac{x(1-x)(1-w+w(cx+d(1-x)))}{x(1-w+w(ax+b(1-x)))+(1-x)(1-w+w(cx+d(1-x)))}.
\end{equation}
We term $\psi(x)$ as the \emph{transient landscape}, which will be shown to be
of great importance in characterizing both deterministic and stochastic dynamics.
It should be mentioned that our definition of the transient landscape can be extended to more general dynamics with
multiple populations (see appendix B).
An relation between $\psi(x)$ and fixation probabilities in \cite{Roca09PLR, Antal BMB 06} will be given in Eq. (\ref{eq_16}) (Also see appendix D).

For the replicator equation
$$\frac{dx(t)}{dt}=x(1-x)[((a-b)x+b)-((c-d)x+d)],$$
It is easy to check that
\begin{align*}
\dot{\psi}(x(t))
&=\frac{d\psi(x)}{dx}\frac{dx}{dt}\\
&=-\ln\left[\frac{1-w+w((a-b)x+b)}{1-w+w((c-d)x+d)}\right]x(1-x)[((a-b)x+b)-((c-d)x+d)]\\
& \leq 0.
\end{align*}
So $\psi(x)$ has the \emph{Lyapunov property} \cite{thompson86book, Hassan02book}, i.e. the dynamical behavior of the replicator
dynamics can be predicted by this transient landscape.
Note that by our definition of $\psi(x)$, the Lyapunov
property is nearly transparent, even for multi-dimensional
systems (see appendix B).

We classify the transient landscape into three main
generic cases (see Fig. 1):

1) \emph{Uni-well}: $a<c$ and $b>d$. $\psi(x)$ decreases on $[0,
x^*]$ and increases on $[x^*, 1]$. Here $x^*=(d-b)/(a-b-c+d)$
is the only minimal extreme point.

2) \emph{Uni-barrier}: $a>c$ and $b<d$. $\psi(x)$ increases on $[0,
x^*]$ and decreases on $[x^*, 1]$. In this
case, the landscape has two local minimal points at both $x=0$ and $1$,
separated by the peak at $x^*$.

3) \emph{Uphill/Downhill}: $a<c$ and $b<d$ (or $a>c$ and $b>d$). In this case, $\psi(x)$
increases (or decreases) on the whole interval $[0, 1]$. So $x=1$ (or x=0) is the minimal point.

Another seemingly trivial case is when $a=c$ and $b=d$, which is
of limited interest in the deterministic dynamics. However, this neutral case with \emph{flat} landscape becomes important in the context of stochastic dynamics.  An interesting result will be
obtained for this case in connection to diffusion approximation
(Sec. \ref{diff_app}).

Note that the rescaled conditional stationary distribution $\eta(x)$ can be expressed as
$$\eta(x)\propto\exp\{-N\psi(x)\},$$
the landscape $\psi(x)$ visualizes the transient dynamics:
The transient system should spend a majority of time around local minimal point(s) in
the landscape.
So we term the minimal point(s) as the \emph{transient attractor(s)}.
In the literature of physics, the transient attractor(s) show
the properties of \emph{metastablity} \cite{Eyring35JCP, Kramers49Phys, Dayue96}. That is, although the ``downhill movement''
towards the local minimum in the landscape maintains the stability of the attractor,
the ``uphill movement'' of crossing the barrier will drive the system to move from the local attractor
to another on a larger time scale. With this observation, we will discuss the fixation from the viewpoint
of the transient landscape.

\subsection{The transient landscape and fixation}

It is known that the process to fixation is intimately dependent upon both the
transient movements before absorption and the one last step to fixation.
Thus we have two cases: First, the
fixation is an inherent result derived directly from the transient
process. Second, the fixation shows distinctly different behavior
from the transient process, i.e., the final fixation does not end
up with attractive absorbing, but one last ``unnatural'' step to
fixation.

Based on the different generic cases of the transient landscape, it
is found that in the \emph{uphill} or \emph{downhill} case, one of the two
absorbing states is located at the transient attractor, we term this
kind of absorbing state as the \emph{attractive absorbing state};
the other absorbing state is called as the \emph{rare absorbing
state}. This classification of the absorbing states is directly linked to
the work by Antal et al. \cite{Antal BMB 06}. We now
denote the probability of fixation at $N$, before reaching
$0$ and starting from the initial state $i$, by $\rho_i$.
Similarly, fixation probability at $0$ starting from $i$ is
denoted by $\gamma_i$.  The explicit expression of $\rho_i$,
for example, can be derived from the following difference
 equation \cite{Traulsen's review}:
$$\rho_j=T_j^-\rho_{j-1}+(1-T_j^--T_j^+)\rho_j+T_j^+\rho_{j+1}$$
with two boundary conditions
$$\rho_0=0,~~~\rho_N=1.$$
Then we have
\begin{equation}
\rho_i=\frac{1+\sum_{k=1}^{i-1}\prod_{j=1}^k\lambda_j}{1+\sum_{k=1}^{N-1}\prod_{j=1}^k\lambda_j},
\end{equation}
where $\lambda_j=T_j^-/T_j^+$.
When $N$ is sufficiently large \cite{Roca09PLR, Antal BMB 06},
\begin{eqnarray*}
\rho_i &=& \frac{1+\sum_{k=1}^{i-1}\prod_{j=1}^kT_j^-/T_j^+}{1+\sum_{k=1}^{N-1}\prod_{j=1}^kT_j^-/T_j^+}
\\
&\approx& \frac{1+\sum_{k=1}^{i-1}e^{N\psi(k/N)}}{1+\sum_{k=1}^{N-1}e^{N\psi(k/N)}}
\\
&\approx& \frac{1+N\int_0^{x} e^{N\psi(y)}dy}{1+N\int_0^{1}e^{N\psi(y)}dy}
\\
&\approx& \frac{\int_0^{x} e^{N\psi(y)}dy}{\int_0^{1}e^{N\psi(y)}dy}.
\end{eqnarray*}
where $x=i/N$.  Therefore,
\begin{equation}
 \lim_{N\rightarrow\infty}
          \frac{1}{N} \ln \frac{d}{dx}\rho_{Nx}
        	= \lim_{N\rightarrow\infty}
 			\frac{1}{N} \ln \frac{d}{dx}\gamma_{N(1-x)}
	    	= \psi(x).
\label{eq_16}
\end{equation}
Eq. (\ref{eq_16}) establishes a connection between our
the transient landscape with fixation probability.
As pointed out by \cite{Antal BMB 06}, in the
\emph{downhill} case that $a>c$ and $b>d$,
$$\rho_1\approx 1-\frac{d}{b},$$
$$\gamma_{N-1}\sim \lambda^N,$$
where
$$\lambda=\frac{d}{b}\frac{(\frac{c}{d})^{(c/(c-d))}}{(\frac{a}{b})^{(a/(a-b)))}}<1.$$
This result shows that the uphill fixation from $N-1$ to $0$ is a
rare event with exponentially small probability, while the downhill
fixation from $0$ to $N-1$ is a rather easy trip. This corresponds
to our ``rare'' or ``attractive'' definition of the absorbing
states. Similarly, in the \emph{uni-barrier} case, both $x=0$ and $1$
are the attractive absorbing states, whereas the barrier
crossing probability from each side to another is exponentially
small.

In the \emph{uni-well} case, however, the only transient attractor is
located at the mixed state $x^*$. In this case, the fixation is not an immediate
result of the transient attraction. Antal et al. \cite{Antal BMB 06} shows
that the fixation time in this case is exponentially large with population size $N$;
while in the other two cases the fixation times have the same approximated
order $N\ln N$. This result is also completely in line with our
classification of the fixation.

The mismatch between the mixed transient attractor and the final
absorbing fixation leads to multiple time scales issue in the
process of evolution. Comparative studies of the mean first passage time
to the attractor $x^*$ and the fixation time
have been carried out in \cite{Ficici JTB 07, Zhou Da JTB 10},
showing the separation of the transient attractive time scale and
the fixation time scale.

Multiple time scales issue is of great importance in
understanding evolutionary systems \cite{Hastings 04}, especially in
explaining the coexistence and extinction of species in ecological
systems \cite{Huisman Nature 99, Huisman Ecology 01}. It has
been reported that the relevant time scale to explain the
coexistence of species in plankton \cite{Huisman Nature 99} is found
in the short term (within a single season in their models). The time
until species being extinct can be much longer than a single season.
Accordingly, the coexistence can be explained here as a
transient phenomenon. The mixed transient attractor, as the stable
equilibrium in the transient dynamics, should be more relevant
within a reasonable time scale. To realize the final fixation, the system has to
escape from the attractor through going uphill on the
landscape, collecting many unfavorable moves consecutively,
for an extremely long time.

\subsection{Stochastic bistability and Maxwell-type construction}

Bistability (or multistability) is one of
the most interesting phenomena in the nonlinear systems
\cite{Gardiner book, Hassan02book, vellela09 JRSI}.
For example, consider the replicator dynamics
$$\frac{dx(t)}{dt}=x(1-x)[((a-b)x+b)-((c-d)x+d)],$$
bistability arises when $a>c$ and $b<d$. In this case,
$x=0$ and $1$ are both stable, separated by the unstable fixed point
$x^*=(d-b)/(a-b-c+d)$. Therefore, the characterizations of
the bistability in the deterministic nonlinear systems should contain
two things: One is \emph{where the attractors are}, the other is
the \emph{basins of attraction}.

One major problem in evolutionary game theory is the selection of
multiple evolutionary stable strategies \cite{szabo2007evolutionary}.
In the bistability case of the deterministic dynamics, the limiting behavior is
determined by its initial state. That is, the measurement of the stability is
closely dependent on the basins of attraction. The stable point with the larger basin of
attraction can be seen as the \emph{risk-dominant} strategy.

In the context of stochastic evolutionary game dynamics, we can also discuss the
noise-induced bistable phenomenon \cite{vellela09 JRSI, qian2009stochastic}.
The bistability in the replicator dynamics corresponds to the \emph{uni-barrier} case
in the transient landscape, where both $x=0$ and $1$ are the local minimal points
in this landscape, separated by the barrier $x^*$. We term this case with two
transient attractors as the \emph{stochastic bistability}.
Furthermore, not only does the landscape
cover the characterizations of the bistability in the replicator dynamics, but we can also give
a straightforward comparison to these two stable states based on this landscape.
From Eq. (\ref{M-potential})
$$\psi(x)=-\int_0^x\ln\left[\frac{1-w+w((a-b)y+b)}{1-w+w((c-d)y+d)}\right]dy,$$
so
$$\psi(0)=0,$$
and
\begin{align*}
\psi(1)&=-\int_0^1\ln\left[\frac{1-w+w((a-b)y+b)}{1-w+w((c-d)y+d)}\right]dy\\
&= \int_0^1\ln(1-w+w((c-d)y+d))dy-\int_0^1\ln(1-w+w((a-b)y+b))dy.
\end{align*}
Without loss of generality, we set $w=1$, then
\begin{align*}
\psi(1)=\psi(0)
& \Longleftrightarrow \int_0^1 \ln\left[ay+b(1-y)\right]dy=\int_0^1\ln\left[cy+d(1-y)\right]dy \\
& \Longleftrightarrow \frac{b\ln b-a\ln a}{b-a}=\frac{d\ln d-c\ln c}{d-c}.
\end{align*}
We term this condition as the \emph{Maxwell-type construction} \cite{ge hao PRL 09}.
Note that
$$\eta(x)\propto\exp\{-N\psi(x)\},$$
so
$$\frac{\eta(0)}{\eta(1)}=\exp\{N(\psi(1)-\psi(0))\}.$$
When $\psi(0)>\psi(1)$,
$$\frac{\eta(0)}{\eta(1)}\rightarrow 0~~as ~N\rightarrow\infty.$$
when  $\psi(0)<\psi(1)$,
$$\frac{\eta(0)}{\eta(1)}\rightarrow \infty~~as ~N\rightarrow\infty.$$
Thereby, even a slight difference between $x=0$ and $1$ in the transient landscape can leads to
a extreme disparity in the distribution (see Fig. 2).
It is observed that except for the critical case, the system will select only one attractor, the global one,
as the unique stable state with the increase of the population size $N$ \cite{ge hao PRL 09, Fudenberg TPB 06}.
In other words, the Maxwell-type construction always singles out the global minimum in the system,
providing another useful criterion for the equilibrium selection.

\section{Diffusion's dilemma of Moran process}

Discrete Markov chain treatment of biological population systems
is necessary for relatively small populations. For large
populations it is convenient and desirable to apply a continuous
approximation \cite{Kimura1955stochastic, Mckane2007singular}.
Beyond the replicator deterministic dynamics as a
continuous limit, a diffusion-type process has long
been much sought after.  However, an important problem
arising is the relation between the original discrete
Markov chain and its approximated representation in term of
a diffusion process \cite{Waxman2011comparison}.  The
perspective of multiple time-scale dynamics in the
previous section provides a better understanding
of this important problem.

The insights we gained
from the transient descriptions leads naturally to a
comparative study of the original discrete-state Moran
process and its continuous-path counterpart.

\subsection{The Kramers-Moyal expansion and landscape
via diffusion approximation}

It is known from the Kramers-Moyal diffusion theory in
physics \cite{Gardiner book} that the Moran process for
large population size can be approximated by a stochastic differential equation. Moran process is a
discrete-time, discrete-state Markov process; its
Kolmogorov forward equation (sometimes called Master
equation) has the form:
\begin{equation}
P_{t+1}(i)-P_t(i) = P_{t}(i-1)T_{i-1}^{+}+ P_t(i+1)T_{i+1}^-
-P_t(i)T_{i}^--P_t(i)T_{i}^+.
\label{master equ}
\end{equation}
When $N$ is large, we take the scalings $x=i/N$, $t'=t/N$,
and the probability density
is $f(x, t)=NP_t(i)$ (we still write $t'$ as $t$).
By performing the truncated Kramers-Moyal (KM) expansion of Eq. (\ref{master equ}),
we have the following approximated Fokker-Planck equation \cite{Traulsen PRL05}:
\begin{equation}
\frac{\partial f(x, t)}{\partial t}=-\frac{\partial}{\partial x}((T^+(x)-T^-(x))f)
+\frac{1}{N}\frac{\partial^2}{\partial x^2}\left(\frac{T^+(x)+T^-(x)}{2}f\right).
\label{FPE}
\end{equation}
This corresponds to the stochastic differential equation
\begin{equation}
dx=(T^+(x)-T^-(x))dt+\sqrt{\frac{T^+(x)+T^-(x)}{N}}dB_t.
\label{SDE}
\end{equation}
where $B_t$ is a Brownian Motion.
In the limit of infinite population size, it is easy to see that Eq. (\ref{SDE}) becomes the deterministic replicator dynamics
\begin{equation}
\dot{x}=T^+(x)-T^-(x).
\end{equation}
In this way, this diffusion approximation links the stochastic Moran process and the macroscopic nonlinear equation.

We should note that the above truncated KM expansion is performed by taking the same
scaling step of time and space with $1/N$. However, when $a=c$ and $b=d$, i.e. the \emph{neutrality} case,
the transient landscape is flat with
$$T^+(x)=T^-(x)$$
for any $x\in(0, 1).$ In this case, the above scaling step is not valid any more.
As a modification, we take $x=i/\sqrt{N}$, then by performing the truncated KM expansion,
we will have
\begin{equation}
\frac{\partial f(x, t)}{\partial t}=\frac{\partial^2}{\partial x^2}\left(\frac{T^+(x)+T^-(x)}{2}f\right).
\end{equation}
This can be well explained by van Kampen's size expansion (see the Appendix C),
which indicates that the scaling of the deterministic drift part should be different
from that of the fluctuated diffusion part | a well-known
fact for the Law of Large Numbers and the Central Limit Theorem.

We now consider the stationary distribution of the diffusion process in Eq. (\ref{FPE}).
The equation satisfied by the stationary distribution should be
\begin{equation}
\frac{\partial}{\partial x}((T^+(x)-T^-(x))\pi(x))
=\frac{1}{N}\frac{\partial^2}{\partial x^2}\left(\frac{T^+(x)+T^-(x)}{2}\pi(x)\right).
\end{equation}
If the boundary conditions are reflecting, the stationary distribution
can be given by
\begin{equation}
\pi(x) \propto \exp (-N \phi(x)),
\label{d-stationary}
\end{equation}
where
\begin{equation}
\phi(x)=-\int_0^x 2\left[ \frac{T^+(y)-T^-(y)}{T^+(y)+T^-(y)} \right]dy.
\label{d-potential}
\end{equation}
However, for the Moran process with absorbing boundaries $T^{+}_0=0$ and $T^{-}_N=0$,
the diffusion approximation should
have corresponding absorbing boundaries \cite{Gardiner book, Mckane2007singular}:
$$f(0,t)=0,~f(1,t)=0.$$
In this way, the stationary distribution in Eq. (\ref{d-stationary}) is not
a real final limiting, but a transient description of the diffusion process.
We term (\ref{d-potential}) as \emph{diffusive landscape}.

\subsection{The validation of KM diffusion approximation in local dynamics}
\label{diff_app}

We now discuss the validation of KM diffusion by comparing the transient landscape
$\psi(x)$ and the diffusive landscape $\phi(x)$.

Consider the derivation of $\psi(x)$, without loss of generality we set $w=1$, then
$$\frac{d\psi(x)}{dx}=-\ln\left[\frac{T^+(x)}{T^-(x)}\right]=-\ln\left[\frac{(a-b)x+b}{(c-d)x+d}\right]=0,$$
where $x^*=(d-b)/(a-b-c+d)$ is the only solution, and
$$\psi^{''}(x^*)=\frac{(a-b-c+d)^2}{bc-ad}.$$
$x^*$ is stable when $\psi^{''}(x^*)>0$, then near $x^*$ we have
$$\psi(x)\approx \psi(x^*)+\psi^{''}(x^*)\frac{(x-x^*)^2}{2}.$$
Meanwhile,
$$\frac{1}{2}\frac{d\phi(x)}{dx}=-\frac{T^+(x)-T^-(x)}{T^+(x)+T^-(x)}=-\frac{((a-b-c+d)x+(b-d)}{(a-b+c-d)x+(b+d)}=0,$$
its only solution is also $x^{*}=(d-b)/(a-b-c+d)$, and interestingly
$$\phi^{''}(x^*)=\frac{(a-b-c+d)^2}{bc-ad}=\psi^{''}(x^*).$$

From the above comparison, we find that $\psi(x)$ and $\phi(x)$ share the same extremal point
and the curvature near $x^*$. That is, the Gaussian variance of $\psi(x)$ is equal to
that of $\phi(x)$,
implying that the local movements near the extremal point in the diffusion
process are in agreement with that in the original Moran process for large populations.

van Kampen's expansion gives a formal argument to the local validation of
diffusion approximation. Consider the VK diffusion (\ref{VK}) near $x^*$,
then Eq. (\ref{VK}) reduces to a time-homogeneous Fokker-Planck equation
\begin{equation}
\frac{\partial \Pi}{\partial t}=-\left(\frac{d}{dx}(T^+(x)-T^-(x))\right)_{x=x*}
\frac{\partial}{\partial \xi}(\xi\Pi)
+\left(\frac{T^+(x^*)+T^-(x^*)}{2}\right)\frac{\partial^2\Pi}{\partial \xi^2}.
\label{linear VK}
\end{equation}
The Gaussian process defined by (\ref{linear VK}) yields to the following linear
stochastic differential equation
$$d\xi(t)=-A\xi(t)dt+DdB_t,$$
where
both $A=-\frac{d}{dx}(T^+(x)-T^-(x))_{x=x*}$ and $D=\sqrt{T^+(x^*)+T^-(x^*)}$ are constant.
This process $\xi(t)$ is called Ornstein-Uhlenbeck (OU) process \cite{Durrett book}, whose
stationary variance is given by
$$ Var(\xi(t))=\frac{D^2}{2A}=\frac{bc-ad}{(a-b-c+d)^2}=\frac{1}{\psi^{''}(x^*)}.$$
This is accordance with our result that diffusion approximation gives the same
local dynamics as the original Moran process for large populations.

We now realize that not only does KM diffusion theory give the deterministic nonlinear dynamical
approximation to the Moran process, but it also gives a good approximation to the intra-attractoral
stochastic dynamics.

\subsection{The invalidation of KM diffusion approximation in global dynamics}

Until now, it has been shown that the KM diffusion approximation correctly describe two kinds of dynamics:
1) Deterministic nonlinear dynamics;
2) local stochastic dynamics.
In this section, we will further investigate the diffusion approximation for global dynamics.

Consider the \emph{uni-barrier} case when $a>c$ and $b<d$. In this bistable game system, the comparison
of different stable strategies is intimately related to the Maxwell-type construction,
which is dependent on the global inter-attractoral dynamics. The Maxwell-type construction
indicates that except for the critical case, only one strategy should be selected as the unique stable one.
Therefore, different constructions could lead to different global dynamical behavior.

For $\psi(x)$, it has been shown that
\begin{align*}
\psi(1)=\psi(0)\Leftrightarrow \frac{b\ln b-a\ln a}{b-a}=\frac{d\ln d-c\ln c}{d-c}.
\end{align*}
For the diffusive landscape $\phi(x)$, however,
\begin{align*}
\phi(1)=\phi(0)
& \Longleftrightarrow \int_0^1 2\left[ \frac{(a-b-c+d)y+b-d}{(a-b+c-d)y+b+d} \right]dy=0\\
& \Longleftrightarrow 2\ln\left[\frac{a+c}{b+d}\right] = \frac{(a-b-c+d)(a-b+c-d)}{(a-b)d-(c-d)b}.
\end{align*}
Fig. 3 shows a simple example:
with the payoff parameters that make
$\psi(1)=\psi(0)$, $\phi(1)>\phi(0)$.
In this case,
the original Moran process and its diffusion approximated process
will select different transient attractors in large population size. Different
global minimum searches lead to different strategy selections.
Therefore, for the evolutionary game systems with multiple stable equilibria,
the validity of this diffusion approximation becomes questionable in global dynamics.

In fact, it is not very surprised to see the global dynamical inconsistency
between the Moran process for large populations and KM diffusion,
since their different large deviation functions result in different
exponential tails of their stationary distributions, which is intimately related to
the inter-attractoral dynamics consisting of barrier crossing movements
from one attractor to another.

To illustrate this problem,
we consider a simple birth-death process $Y(t)$ with birth rate $\mu_i=\mu$ and death rate
$\lambda_i=\lambda$, i.e. the transition rates are independent of the states \cite{ qian_2011_nonlinearity_review}.
$Y(t)$ has a reflecting boundary at $M$ and an absorbing boundary at $0$. We are interested in
$\tau_n$ the first passage time from $n$ to $0$
\cite{Gardiner book}. In this simple model, there are three
kinds of movements from $n$ to $0$ (let $\theta=\lambda/\mu$): downhill ($\theta>1$), uphill ($\theta<1$) and
flat ($\theta=1$).

It is not difficult to have that
\begin{equation}
	\tau_n = \frac{1}{\mu-\lambda}\left(\frac{1-\theta^n}
		{\theta^n-\theta^{n+1}}\right)
		+\frac{n}{\lambda-\mu}.
\label{tau_n}
\end{equation}
Let the space step between $i$ to $i+1$ be $\delta$, and
let $\delta\rightarrow 0$ and $n\rightarrow\infty$, but $n\delta\rightarrow x$,
then we have a Fokker-Planck equation
\begin{equation}
	\frac{\partial f(x,t)}{\partial t} = D\frac{\partial^2 f}{\partial x^2}
			- V\frac{\partial f}{\partial x},
\label{FP}
\end{equation}
where $D=(\mu+\lambda)\delta^2/2$ and $V=(\mu-\lambda)\delta$.  The
corresponding first passage time for (\ref{FP}) is
\begin{equation}
	\widetilde{\tau_x} = \frac{1}{V}\left[\frac{D}{V}\left(e^{\frac{V}{D}x}-1\right)-x\right].
\end{equation}
Now discretizing $x$ as $n\delta$, we have
\begin{equation}
	\widetilde{\tau}_n = \frac{1}{(\mu-\lambda)}\frac{1+\theta}{2(1-\theta)}
		\left(e^{\frac{2(1-\theta)n}{(1+\theta)}}-1\right)
				+\frac{n}{\lambda-\mu}.
\label{tau_n 2}
\end{equation}
Comparing $\tau_n$ and $\widetilde{\tau}_n$
\begin{equation}
	\lim_{n\rightarrow\infty}\frac{\tau_n}{\widetilde{\tau}_n} = \lim_{n\rightarrow\infty}\frac{\left(\frac{1-\theta^n}
		{\theta^n-\theta^{n+1}}\right)-n}
			{\frac{(1+\theta)}{2(1-\theta)}
		\left(e^{\frac{2(1-\theta)n}{(1+\theta)}}-1\right)
				-n }=\left\{\begin{array}{cl}  \infty &~~if~ \theta <1  \\
					1 &~~ if~\theta\ge 1
				\end{array}\right.
\end{equation}
More specifically, we have
\begin{equation}
	\lim_{n\rightarrow\infty}
	-\frac{1}{n}\ln\frac{\tau_n}{\widetilde{\tau}_n} =
		\frac{2(1-\theta)}{1+\theta}+\ln\theta,  \ \ \ ( ~\theta<1)
\end{equation}
and
\begin{equation}
	\lim_{n\rightarrow\infty}
	 n\ln\frac{\tau_n}{\widetilde{\tau}_n} = \left\{ \begin{array}{cl}
			1 & if~\theta = 1 \\[7pt]
		\frac{1}{2} & if~\theta>1
		\end{array}\right.
\end{equation}

From the above comparison of $\tau_n$ and $\widetilde{\tau}_n$,
we find that they both approach to $n/(\lambda-\mu)$ in the downhill dynamics; while
in the uphill dynamics,
both $\tau_n$ and $\widetilde{\tau}_n$ share the exponential form of $\sim e^{\alpha n}$,
but different exponential parameters. This is the heart of our example. We should note that,
for the bistable systems, the Maxwell-type constructions are determined by the jump processes
between these two attractors (back and forth), which are both rare events with exponentially long time to happen.
According to the above disparity between  $\tau_n$ and $\widetilde{\tau}_n$ in the uphill dynamics, KM diffusion
approximation can not give the exponent correctly, and then
results in representing the inter-attractoral dynamical
inaccurately. We suggest this as the reason for the
invalidity of the diffusion approximation for the global
dynamics and landscape.

\subsection{Diffusion's dilemma}

According to Kurtz's theorem \cite{Kurtz1976, Kurtz1978}, KM's diffusion theory can be mathematically justified
only for any finite time $t$. In other words, Eq. (\ref{FPE}) correctly approximates the finite-time Moran process
for large but finite populations, whereas it is not guaranteed
that they share the same long-term stationary behavior.
Therefore, the difficulty encountered by KM's diffusion
in bistable game systems stems from the fact that exchanging
the limits of population size and time is problematic.
It concerns with non-uniform convergence of Kurtz's result.

A natural question is whether one can find a diffusion
process that gives both satisfactory finite-time and
stationary dynamical approximation.
H\"{a}nggi et al. \cite{hanggi1984bistable} proposed a
very different diffusion process in the context of
Chemical Master Equation:
\begin{equation}
\frac{\partial f(x, t)}{\partial t}=-\frac{\partial}{\partial x}((T^+(x)-T^-(x))f)
+\frac{1}{N}\frac{\partial^2}{\partial x^2}\left(\frac{T^+(x)-T^-(x)}{\ln T^+(x)-\ln T^-(x)}f\right).
\label{FPE-2}
\end{equation}
The heuristic derivation of Eq. (\ref{FPE-2})
is based on Onsager's theory. Then the stochastic potential for the system
should be the transient landscape $\psi(x)$, and the \emph{thermodynamic force} is
$$F(x)=-\frac{d\psi(x)}{dx}=\ln T^+(x)-\ln T^-(x).$$
Therefore, the macroscopic ordinary differential
equation should be
\begin{equation}
	\frac{dx}{dt}=T^+(x)-T^-(x)=\eta^{-1}(x) F(x).
\end{equation}
So
\begin{equation}
	\eta^{-1}(x)=\frac{T^+(x)-T^-(x)}{\ln T^+(x)-\ln T^-(x)},
\end{equation}
and the diffusion coefficient proportional to $\eta^{-1}(x)$.
In order to distinguish
H\"{a}nggi et al.'s from KM's, we term Eq. (\ref{FPE-2}) as HGTT's diffusion.

It is easy to show that Eq. (\ref{FPE-2}) gives the same large deviation function
as the original Moran process. Moreover, by comparing the drift coefficients of
(\ref{FPE}) and (\ref{FPE-2}),
$$a_{KM}(x)=a_{HGTT}(x)=T^+(x)-T^-(x),$$
HGTT's and KM's describe the same ODE when $N$ tends to infinity.
For the diffusion coefficients:
$$b_{KM}(x)=\frac{T^+(x)+T^-(x)}{2},$$
$$b_{HGTT}(x)=\frac{T^+(x)-T^-(x)}{\ln T^+(x)-\ln T^-(x)}.$$
It is easy to find that HGTT's diffusion
coefficient is always smaller than that of KM's (see Fig. 4),
except $b_{KM}(x) \approx b_{HGTT}(x)$ when $x$ near $x^*$.
So away from the extremal point, HGTT's diffusion shows different
finite-time stochastic dynamics from KM's.
Note that KM's diffusion
gives the correct finite-time dynamical approximation of the original Moran process,
HGTT's could then show a wrong short-term dynamics for most of the initial states.

Therefore, our diffusion dilemma can be
stated as follows: Can we find an approximated diffusion process correctly describe the whole three dynamical regimes:
(a) The deterministic limit;
(b) the short time stochastic dynamics;
(c) long time global dynamics?
For truncated KM approximation (and van Kampen's expansion), the
(a) and (b) are correct for each and very attractor, but (c) is wrong.
For HGTT's diffusion, (a) and (c) are correct, but (b) is
wrong. So we can not find a diffusion process that provides all the three correctly.

\section{discussions}

Stochastic dynamics have become a fundamental theory in understanding Darwinian evolutionary theory
\cite{Nowak book, Traulsen's review, sandholm2011population, ao2005laws, Ao08CTP}.
Besides \emph{nonlinearity}, \emph{stochasticity} has been shown
as another basic feature of complexity in biological world \cite{allen2003introduction},
especially within the scale of cellular dynamics
\cite{elowitz2002stochastic, cai2006stochastic, beard2008chemical}.
Stochastic evolutionary game dynamics, as \emph{agent-based} models to describe the
kinetics in polymorphic population systems, offer a framework to study the
\emph{frequency-dependent selection} in evolution.

The present paper discuss the well-mixed stochastic evolutionary game dynamics from the viewpoint of
the \emph{transients}. The transient landscape, as a potential-like representation of the
pre-fixation dynamics, has been constructed via the conditional stationary distribution in the theory of quasi-stationarity in
terms of the large deviation rate function.  The involvement
of large deviation theory from probability is essential here,
for without it, the landscape would be system's size dependent.
It has been shown that this transient landscape can play a central role in connecting the deterministic
replicator dynamics, the final fixation behavior and diffusion approximation. As a Lyapunov function
of the replicator dynamics, the transient landscape visually
captures the infinite-population nonlinear behavior.
The downhill movements in this landscape corresponds to the dynamics of its deterministic counterpart,
whereas the rare uphill movements arising from the random fluctuations are of more interest
in stochastic evolutionary systems. To capture the eventual fixation behavior from the transient perspective,
we have classified the absorbing states into two cases: The attractive absorbing state which is located
at the transient attractor; the other rare absorbing state which is located at the top of the landscape.
The former is an inherent result of the transient downhill dynamics, while the latter is related to the
multiple time scale issue, that is, the final fixation time scale is separated from the transient coexistence
quasi-stationarity.

Furthermore,
the Maxwell-type construction and diffusion approximation are both important problems linking to the transient dynamics.
The Maxwell-type construction is a global description of nonlinear bistable stochastic dynamics,
which is not present in deterministic dynamics. This construction always searches the global minimum in the landscape,
so it is a direct result of inter-attractoral dynamics. The comparison of the Maxwell-type constructions between the original
transient landscape and its diffusion counterpart indicates that the truncated KM diffusion approximation could result in
different global dynamics, that is, the original Moran process for large populations and its diffusion counterpart
could select different global stable points.
In order to solve this problem, another HGTT's diffusion has been constructed for giving the correct
long-term asymptotic dynamics. However, this diffusion gives the wrong finite time stochastic dynamics.

By investigating the first passage times in the simple birth-death process,
it has been found that the failure of exponential approximation in the uphill movement could be a reason for our diffusion's problem.
Mathematically, the diffusion approximation is just a second-order polynomial expansion of the Master equation,
which only offers the second-order precision for the original process. Accordingly, this approach can give the correct
deterministic dynamics (first order) and Gaussian dynamics near the stable point (second-order). However,
the inter-attractoral dynamics is determined by the rare barrier crossing
movements with exponentially small probabilities, so the Maxwell-type construction should be approximated
in the level of exponential asymptotics, which could be out of any finite order expansions' league. In the theory of
probability, this is the domain of the Large Deviation Theory \cite{Dembo}.

It is believed that discrete stochastic dynamics offers a new perspective on biological dynamics.
Besides the conventional concentrations on
maximum-likelihood events, more attention should be paid to \emph{rare events}. Evolution itself
is a process with the accumulations of various rare events, such as genetic or epigenetic mutations
and ecological catastrophes. So the stochasticity is not just fluctuations near the most probable
macroscopic states, but an important source of complexity,
i.e., ``innovation'', especially on an evolutionary time scale.

\section{acknowledgementS}

We thank Tibor Antal, Ping Ao and Hao Ge for reading the manuscript and helpful comments.
Discussions with Jiazeng Wang, Bin Wu and Michael Q. Zhang are gratefully
acknowledged.
DZ also wish to acknowledge support by the National Natural Science Foundation
of China (10625101), and the 973 Fund (2006CB805900).

\section{Appendix}

\subsection{Conditional stationary distribution in the theory of quasi-stationarity}
\label{appendix A}

Quasi-stationarity is a series of stochastic mathematical techniques
for analyzing the Markov processes with absorbing states. The basic idea
of the quasi-stationarity is to find some \emph{effective} distributions
for characterizing the transient behavior of the process. There are
basically two kinds of quasi-stationarities: conditional stationary distribution
and stationary conditional one. Here we only consider
the former, see \cite{Dorroch 1965} for more details.

In order to introduce the conditional stationary distribution,
we now add small mutations to the original Moran process as follows:
 \begin{equation}
T_0^+=P(0\rightarrow 1)=\varepsilon.
\end{equation}
\begin{equation}
T_N^-=P(N\rightarrow N-1)=\varepsilon.
\end{equation}
In this case, the process has become irreducible.
Further, the stationary distribution of the new chain reads:
\begin{equation}
\mu_i(\varepsilon)=C\prod_{k=1}^iT_{k-1}^+/T_k^-.
\end{equation}
where $C$ is the normalized constant.
Consider
\begin{equation}
\eta_i(\varepsilon)=\frac{\mu_i(\varepsilon)}{1-\mu_0(\varepsilon)-\mu_N(\varepsilon)}~~~~~~i=1,
2, \ldots, N-1,
\end{equation}
it is not difficult to have that $\eta(\varepsilon)$ is independent of
$\varepsilon$, and $\eta(\varepsilon)$ is just the same as $\eta$ in Eq. (\ref{CSD}).
It has been shown in \cite{Dorroch 1965} that $\eta_j$ is proportional to the expected time of visits to state $j$ before absorption when started in the revival distribution. That is, $\eta$ characterizes the occupation time distribution of the transient dynamics. The larger $\eta_j$, the longer the process stays at state $j$ before absorption.

Here we should emphasize that, given a Markov chain with absorbing states, the pre-fixation occupation
time distribution depends on the distribution of the states in which the chain is revived.
For the birth-death process here, it is natural to choose the reviving states as
neighboring the absorbing states.

\subsection{Generalized transient landscape for multi-dimensional cases}
\label{appendix B}

In this section we will show that the definition of the \emph{transient landscape} in Eq. (\ref{M-potential})
can be extended to more general cases.

Consider a multi-dimensional birth-death process with absorbing states, i.e. $X_t=(X_1(t), X_2(t), ..., X_d(t))$. The state space of
this process is a $d$-dimensional vector space, denoted as $\mathbf{N}^d$. In the generalized Moran process \cite{Traulsen06PRE},
for instance, $d$ is the number of strategies, and $X_i(t)$ is the number of individuals with strategy $i$ at time $t$.

Suppose $X_t$ has a unique conditional stationary distribution $P^{N}_{ss}(\vec{n})$, where $N$ is the population size,
$\vec{n}\in\mathbf{N}^d$. As a function of $N$, $P^{N}_{ss}(\vec{n})$ usually has the so called WKB expansion \cite{qian_2011_nonlinearity_review}
for large population size:
$$P^{N}_{ss}(\vec{n})\propto \exp\left[-N\varphi(\vec{x})+\varphi_1(\vec{x})+\frac{\varphi_2(\vec{x})}{N}+...\right]. ~~~\vec{x}=\frac{\vec{n}}{N}.$$
That is, $\varphi(\vec{x})$ can be obtained from
$$\varphi(\vec{x})=\lim_{N\rightarrow\infty}-\frac{1}{N}\ln P^{N}_{ss}(N\vec{x}),$$
if the above limit exists. We define $\varphi(\vec{x})$ as the generalized transient landscape.

It has been shown that $\varphi(\vec{x})$ still has the Lyapunov property with respect to its thermodynamic
limit \cite{Hu1986}. Suppose the thermodynamic limit of $X_t$ can be described as the following deterministic
differential equations
$$\frac{d\vec{x_t}}{dt}= \vec{a}(\vec{x}_t), ~~~\vec{x}_t={X_t}/N.$$
In particular, for the generalized Moran process \cite{Traulsen06PRE},
$$\vec{a}(\vec{x})=(..., a_i(\vec{x}),...)^T=
(..., \sum_{j=1}^{d}(T_{ji}(\vec{x})-T_{ij}(\vec{x})),...)^T,$$
where $T_{ij}(x)$ is the frequency-dependent probability that an $i$ strategist is replaced by a $j$
strategist.
From \cite{qian_2011_nonlinearity_review, Hu1986}, we have
\begin{align*}
\dot{\varphi}(\vec{x_t})
&=\bigtriangledown\varphi(\vec{x})\cdot\frac{d\vec{x}}{dt}\\
&=\bigtriangledown\varphi(\vec{x})\cdot \vec{a}(\vec{x})\\
&=-(\bigtriangledown\varphi(\vec{x}))^2\\
& \leq 0.
\end{align*}

\subsection{van Kampen's expansion}
\label{appendix C}

van Kampen's expansion provides another systematic method of diffusion approximation \cite{van Kampen}.
The idea of VK expansion is that, in large population size $N$, the number we are interested in
(e.g. the number of strategy $A$) is expected to consist of two parts: deterministic and fluctuations
parts. Consider the continuous time birth-death process here (the discrete time case is similar),
for any state $i$, we have
$$i=Nx(t)+N^{1/2}\xi(t),$$
where $x(t)$ is of order $N^{-1}$, $\xi(t)$ is of $N^{-1/2}$.
Define the shift operators as
$\omega (T_i)=T_{i+1}$ and $\omega^{-1}(T_i)=T_{i-1}$,
so the Master equation can be written as
$$\frac{dP_t(i)}{dt}=(\omega^{-1}-1)(T^+_iP_t(i))+(\omega-1)(T^-_iP_t(i)),$$
where $T^+_i$ is the birth rate, and $T^-_i$ is the death rate.
Now we denote the distribution of $\xi(t)$ as $\Pi(\xi, t)$. In fact,
$$\Pi(\xi, t)=P_{t}(Nx(t)+N^{1/2}\xi(t)),$$
and we have
$$\frac{dP_t(i)}{dt}=\frac{\partial\Pi(\xi, t)}{dt}+\frac{\partial\Pi(\xi, t)}{d\xi}\frac{d\xi}{dt}
=\frac{\partial\Pi(\xi, t)}{dt}-N^{1/2}\frac{\partial\Pi(\xi, t)}{d\xi}\frac{dx(t)}{dt}.$$
We take the Taylor expansions:
$$\omega-1 \approx N^{-1/2}\frac{\partial}{\partial \xi}+\frac{1}{2}N^{-1}\frac{\partial^2}{\partial \xi^2},$$
$$\omega^{-1}-1 \approx -N^{-1/2}\frac{\partial}{\partial \xi}+\frac{1}{2}N^{-1}\frac{\partial^2}{\partial \xi^2},$$
$$T^+_{i} \approx N T^+(x)+N^{1/2} \xi (\frac{dT^+(x)}{dx}),$$
$$T^-_{i} \approx N T^-(x)+N^{1/2} \xi (\frac{dT^-(x)}{dx}),$$
where
$$T^{+}(x)=\lim_{N\rightarrow \infty} \frac{T^{+}_{Nx}}{N}, T^{-}(x)=\lim_{N\rightarrow \infty} \frac{T^{-}_{Nx}}{N}.$$
So
\begin{align*}
\frac{\partial\Pi(\xi, t)}{dt}-N^{1/2}\frac{\partial\Pi(\xi, t)}{d\xi}\frac{dx}{dt}
=\left(-N^{-1/2}\frac{\partial}{\partial \xi}+\frac{1}{2}N^{-1}\frac{\partial^2}{\partial \xi^2}\right)
\left((N T^+(x)+N^{1/2} \xi (\frac{dT^+(x)}{dx})\Pi\right)\\
+\left(N^{-1/2}\frac{\partial}{\partial \xi}+\frac{1}{2}N^{-1}\frac{\partial^2}{\partial \xi^2}\right)
\left((N T^-(x)+N^{1/2} \xi (\frac{dT^-(x)}{dx})\Pi\right).
\end{align*}
The terms of order $N^{1/2}$ on either side will vanish if $x(t)$ satisfies the equation
$$\frac{dx}{dt}=T^+(x)-T^-(x),$$
which is just the deterministic replicator dynamics.
If consider the terms of order $N^{0}$, $\xi(t)$ should obeys
\begin{equation}
\frac{\partial \Pi}{\partial t}=-\left(\frac{d}{dx}(T^+(x)-T^-(x))\right)
\frac{\partial}{\partial \xi}(\xi\Pi)
+\left(\frac{T^+(x)+T^-(x)}{2}\right)\frac{\partial^2\Pi}{\partial \xi^2}.
\label{VK}
\end{equation}
This is a linear Fokker-Planck equation whose coefficients only depend on $x(t)$.
So van Kampen's approach gives the correct dynamics conditioned on the deterministic solution.
If we substitute $z=N^{-1/2}\xi+x(t)$, we can find that Eq. (\ref{VK}) is exactly the same as Eq. (\ref{FPE}).

\subsection{The relations between transient landscape and fixation probability}
\label{appendix D}

Eq. (\ref{eq_16}) shows that the fixation probabilities,
$\rho_j=1-\gamma_j$ and our transient landscape $\psi(x)$
have the following relation:
\begin{equation}
 \lim_{N\rightarrow\infty}
          \frac{1}{N} \ln \frac{d}{dx}\rho_{Nx}
        	= \lim_{N\rightarrow\infty}
 			\frac{1}{N} \ln \frac{d}{dx}\gamma_{N(1-x)}
	    	= \psi(x).
\end{equation}
To further illustrate this, let us
consider a similar relation in a diffusion process with
the following stochastic differential equation
\begin{equation}
dx=a(x)dt+\frac{1}{\sqrt{N}}b(x)dB_t
\label{1}
\end{equation}
with absorbing boundary conditions. As shown in Sec. IV, the conditional stationary distribution
can be obtained by solving the Kolmogorov forward equation
\begin{equation}
-\frac{\partial}{\partial x}\left(a(x)\pi(x)\right)
+\frac{1}{2N}\frac{\partial^2}{\partial x^2}\left(b^2(x)\pi(x)\right)=0,
\end{equation}
where the stationary distribution is
\begin{equation}
\pi(x) \propto \exp (-N \phi(x)),
\label{onedimension}
\end{equation}
and the transient landscape is
\begin{equation}
\phi(x)=-\int_0^x 2\left[ \frac{a(y)}{b^2(y)} \right]dy.
\end{equation}
On the other hand, the fixation probability from $x$ to $1$ is the solution of
the backward equation \cite{Gardiner book}
\begin{equation}
a(x)\frac{\partial}{\partial x}\rho(x)
+\frac{1}{2N}b^2(x)\frac{\partial^2}{\partial x^2}\rho(x)=0,
\end{equation}
with boundary conditions
$$\rho(0)=0,~~\rho(1)=1.$$
It is not difficult to show that
$$\rho(x)=\frac{\int_0^{x} e^{N\phi(y)}dy}{\int_0^{1}e^{N\phi(y)}dy},$$
so we also have
\begin{equation}
 \lim_{N\rightarrow\infty}
          \frac{1}{N} \ln \frac{d}{dx}\rho(x)
	    	= \phi(x).
\label{2}
\end{equation}

We now attempt to generalize the above relation
in Eq. (\ref{2}) to the more general multi-dimensional cases.
Consider an $n$-dimensional diffusion process
with forward equation
\begin{equation}
\frac{\partial f(\vec{x}, t)}{\partial t}=
	\sum_i^{n}\frac{\partial}{\partial x_i}\left\{
             -A_i(\vec{x})f + \frac{1}{2N}\sum_j^n
          \frac{\partial}{\partial x_j}(B_{ij}(\vec{x})f) \right\},
~~\vec{x}\in D,
\label{48}
\end{equation}
where the absorbing boundary of $D$ is denoted as $\partial D$.
For any $a\in \partial D$, the fixation probability density
at $a$ from $x$ also satisfies the backward equation
\begin{equation}
\sum_i^{n}A_i(\vec{x})\frac{\partial}{\partial x_i}\rho_{\vec{x}}(\vec{a})
+\frac{1}{2N}\sum_{i,j}^{n}B_{ij}(\vec{x})\frac{\partial}{\partial x_i}\frac{\partial}{\partial x_j}\rho_{\vec{x}}(\vec{a})=0.
\label{49}
\end{equation}
Its boundary condition is
$$\rho_{\vec{x}}(\vec{a})=\delta_{\vec{x}-\vec{a}},$$
where $\delta_{\vec{x}-\vec{a}}$ is the Dirac-delta function
for $\partial D$.

The conditional stationary distribution solves the $\{\cdots\}=-J_i\left(\vec{x}\right)$ in Eq. (\ref{48}), where
$\nabla\cdot J(\vec{x})=0$.  Detailed balance, however,
further dictates $J_i(\vec{x})\equiv 0$  \cite{Gardiner book}.
Therefore,
\begin{equation}
      -\frac{1}{2}\sum_{j}^nB_{ij}(\vec{x})\frac{\partial}
        {\partial x_j}\phi(\vec{x})  =
\left[ A_i(\vec{x})-\frac{1}{2N}\sum_{j}^n\frac{\partial}{\partial x_j}B_{ij}(\vec{x})\right],
\label{50}
\end{equation}
where $\phi(\vec{x})$ is our transient landscape.

Now consider Eq. (\ref{49}) in the light of (\ref{50}).
First we denote $\vec{\zeta}(\vec{x})
=\nabla_{\vec{x}}\rho_{\vec{x}}(\vec{a})$.  It
satisfies
\begin{eqnarray}
0 &=& \sum_i^{n}A_i(\vec{x})\zeta_i(\vec{x})
+\frac{1}{2N}\sum_{i,j}^{n}B_{ij}(\vec{x})\frac{\partial}{\partial x_j}
\zeta_i(\vec{x})
\nonumber\\
&=&  \frac{1}{2} \sum_{i,j}^{n}
   \left[\frac{1}{N}\frac{\partial}
          {\partial x_j}B_{ij}(\vec{x})
       -B_{ij}(\vec{x})\frac{\partial}
        {\partial x_j}\psi(\vec{x})
\right] \zeta_i(\vec{x})
+\frac{1}{2N}\sum_{i,j}^{n}B_{ij}(\vec{x})\frac{\partial}{\partial x_j}
\zeta_i(\vec{x})
\nonumber\\
&\approx& -\frac{1}{2} \sum_{i,j}^{n}
   \left[B_{ij}(\vec{x})\frac{\partial}
        {\partial x_j}\phi(\vec{x})
\right] \zeta_i(\vec{x})
+\frac{1}{2N}\sum_{i,j}^{n}B_{ij}(\vec{x})\frac{\partial}{\partial x_j}
\zeta_i(\vec{x}).
\nonumber\\
&=& -\frac{1}{2} \sum_{i,j}^{n}
    \zeta_i(\vec{x})B_{ij}(\vec{x})\frac{\partial}
        {\partial x_j} \left[ \phi(\vec{x})
         -\frac{1}{N}\ln \zeta_i(\vec{x}) \right].
\end{eqnarray}
We see a hint of Eq. (\ref{2}) in the square bracket.
For multi-dimensional problems, the gradient of
$\rho_{\vec{x}}(\vec{a})$ is a vector while $\phi(x)$ is a scalar.
Therefore, it seems to us, even with detailed balance condition,
the relation in Eq. (\ref{2}) can not be generalized to
multi-dimensional case.
On the other hand, the definition of $\phi(x)$ can be generalized to multi-dimensional case (see appendix B), even
though finding it will be hard.

\newpage
\section*{Captions}

Figure 1 (Color online): Transient landscapes and conditional stationary distributions:
(a) Uni-well case: The small window shows the transient landscape $\psi(x)$; The large
window shows the conditional stationary distribution $\eta(x)$ with different population size.
Parameters are $a=1$, $b=2$, $c=3$, $d=1$, and $w=0.7$.
(b) Uni-barrier case, with parameters $a=2.5$, $b=1$, $c=1$ ,$d=2$ and $w=0.7$.
(c) Uphill case, with parameters $a=1$, $b=1.2$, $c=1.5$, $d=1.4$.

Figure 2 (Color online): Maxwell-type construction for the bistable Moran process:
(a) When the critical condition is satisfied ($a=2$, $b=1$, $c=1$, $d=2$), $\psi(0)=\psi(1)$.
Both are equally important.
(b) With parameters $a=2.5$, $b=1$, $c=1$, $d=2$, $\psi(0)>\psi(1)$. Then $\eta(0)<\eta(1)$,
and even $\eta(0)\ll\eta(1)$ for large population size.

Figure 3 (Color online): The original Moran process and its KM approximated diffusion process
show different Maxwell-type constructions. In this example,
$\psi(1)=\psi(0)$, but $\phi(1)>\phi(0)$.
(The figure is magnified and we focus on the region near $x=1$)

Figure 4 (Color online): The HGTT's diffusion coefficient is always smaller than KM's, except
at $x^*$ where they are both equal to each other.

\section*{Figures}
\newpage

\begin{figure}
\begin{center}
\subfigure[] {\includegraphics[width=0.6\textwidth]{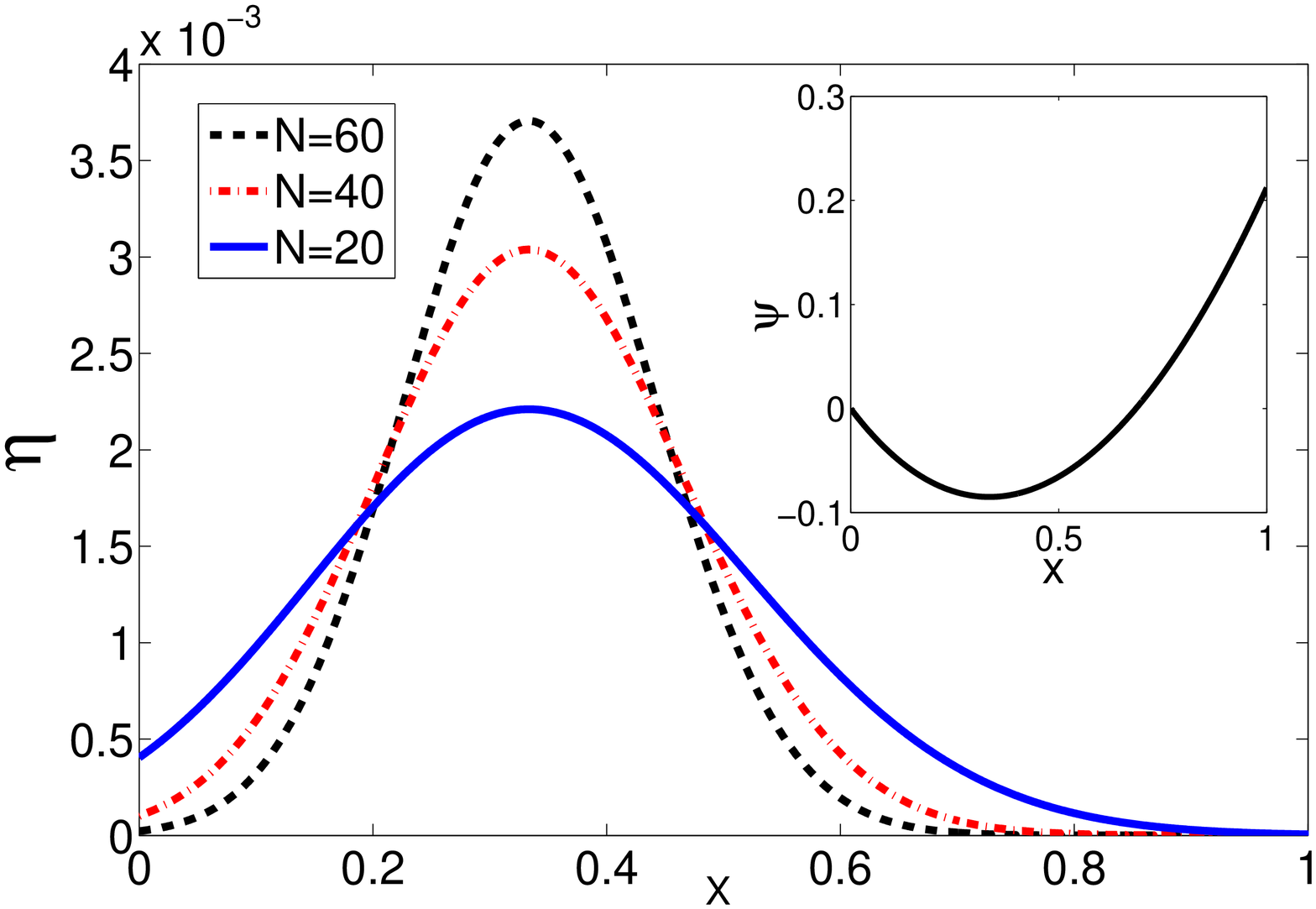}}
\subfigure[] {\includegraphics[width=0.6\textwidth]{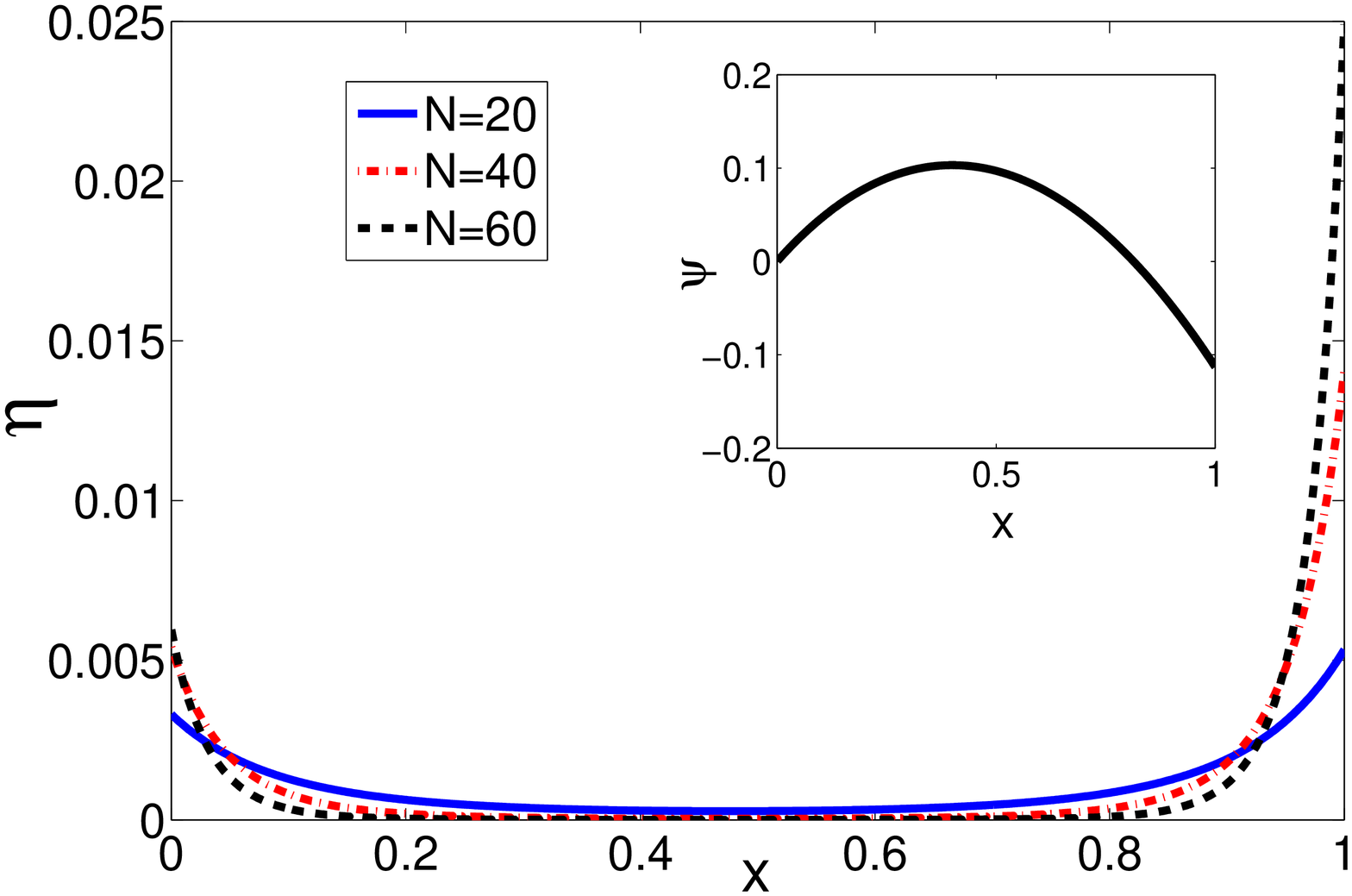}}
\subfigure[] {\includegraphics[width=0.6\textwidth]{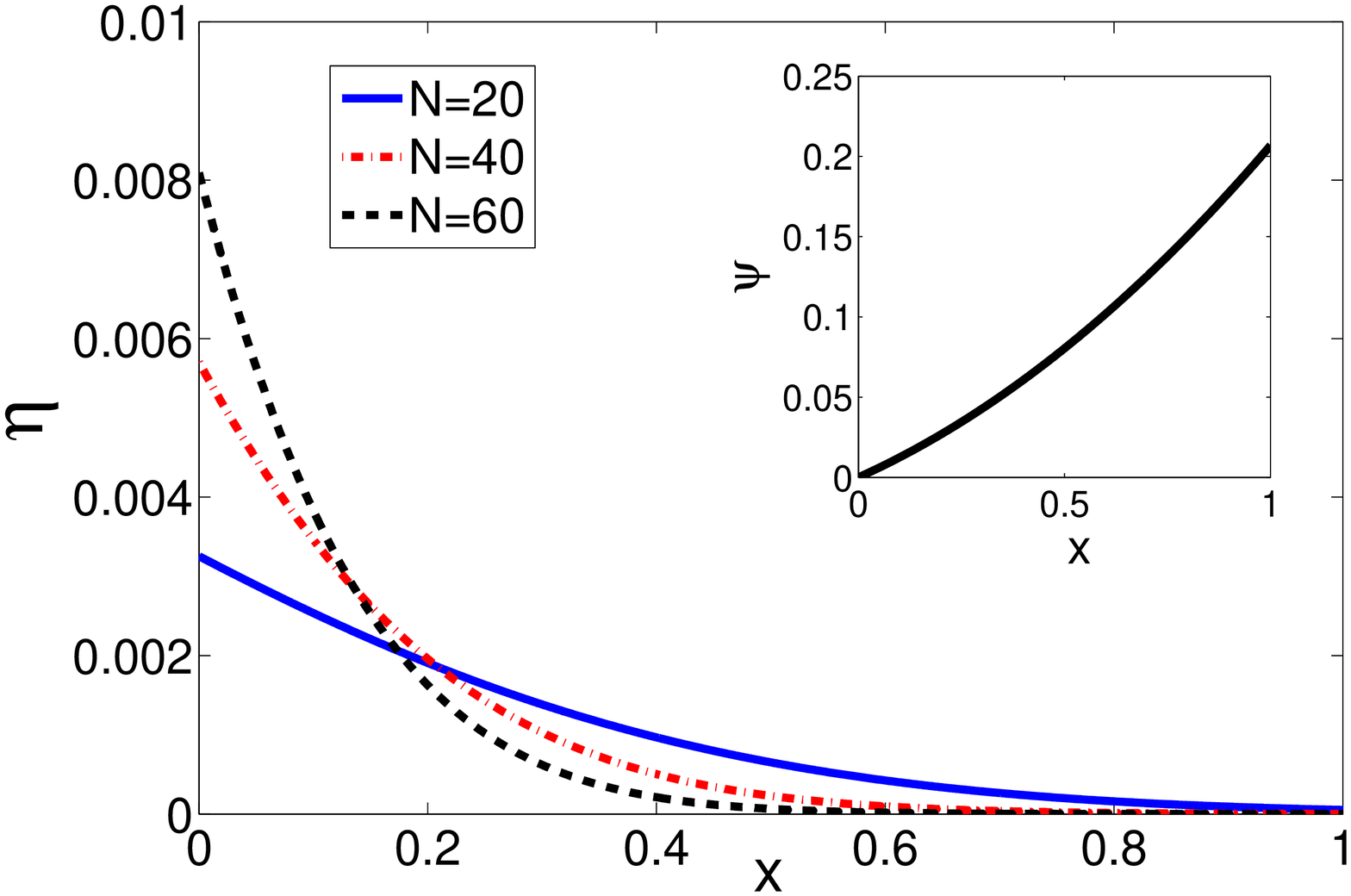}}
\caption{}
\label{FIG_1}
\end{center}
\end{figure}

\begin{figure}
\begin{center}
\subfigure[] {\includegraphics[width=1\textwidth]{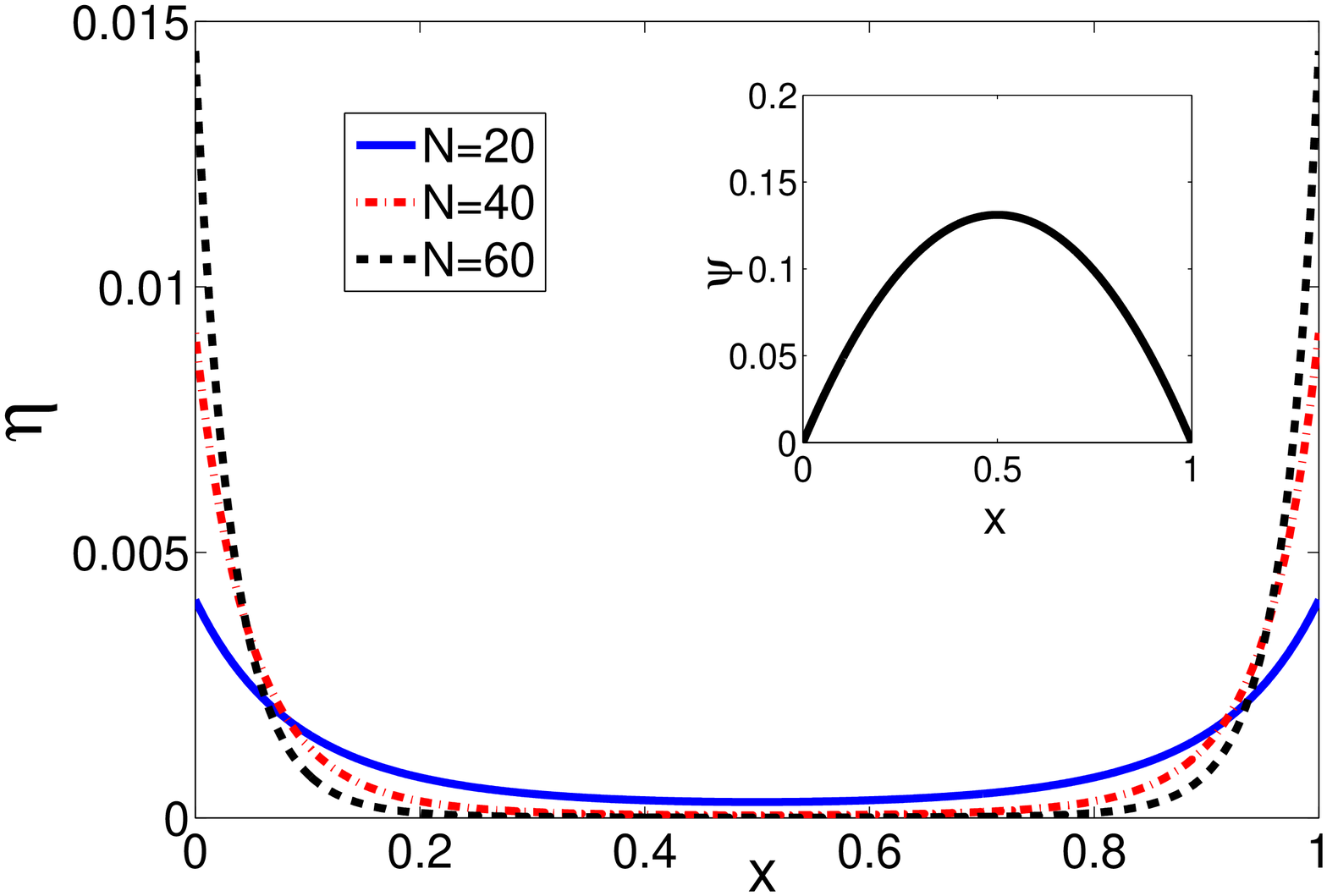}}
\subfigure[] {\includegraphics[width=1\textwidth]{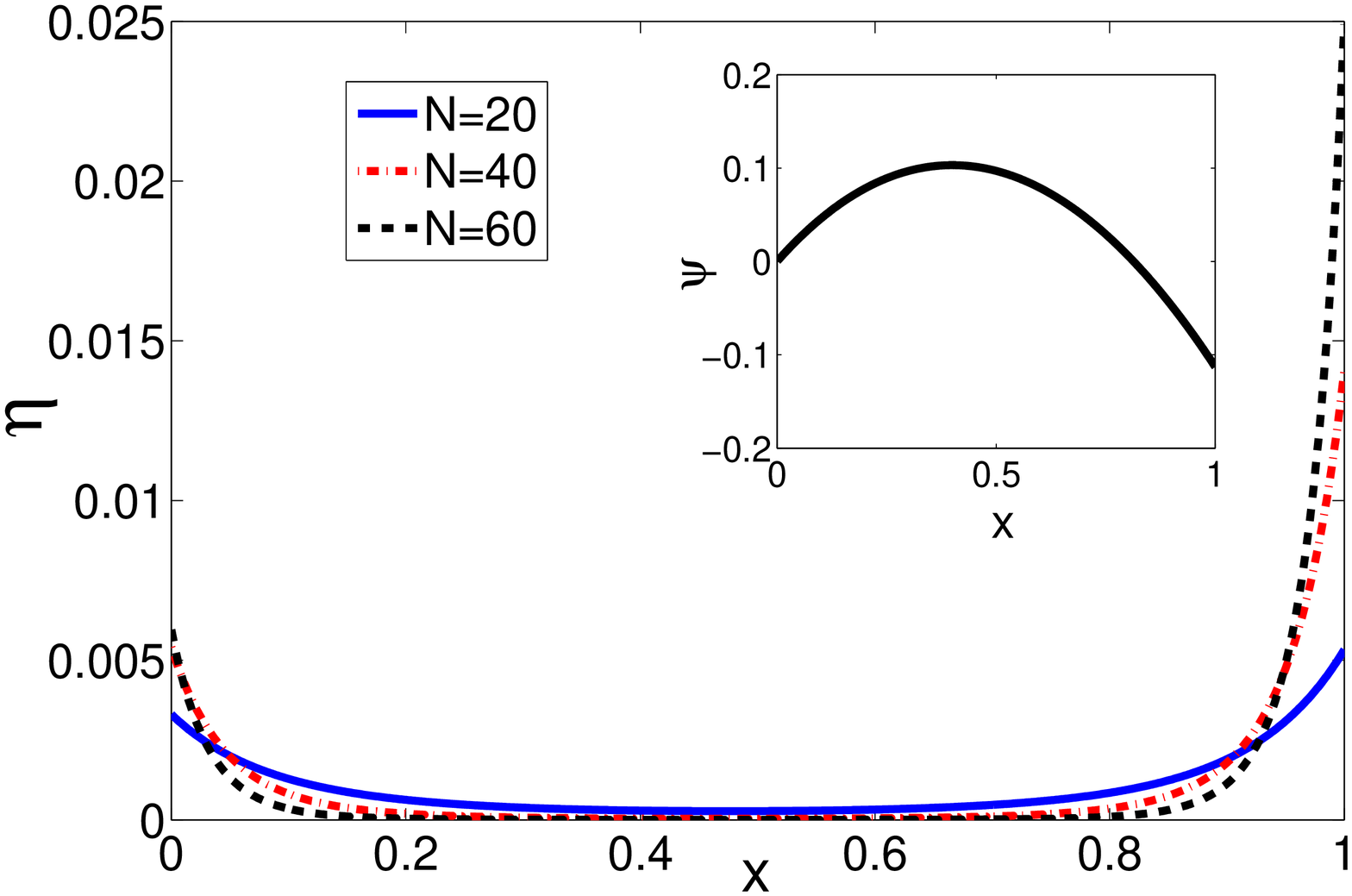}}
\caption{}
\label{FIG_2}
\end{center}
\end{figure}

\begin{figure}
\begin{center}
\includegraphics[width=1\textwidth]{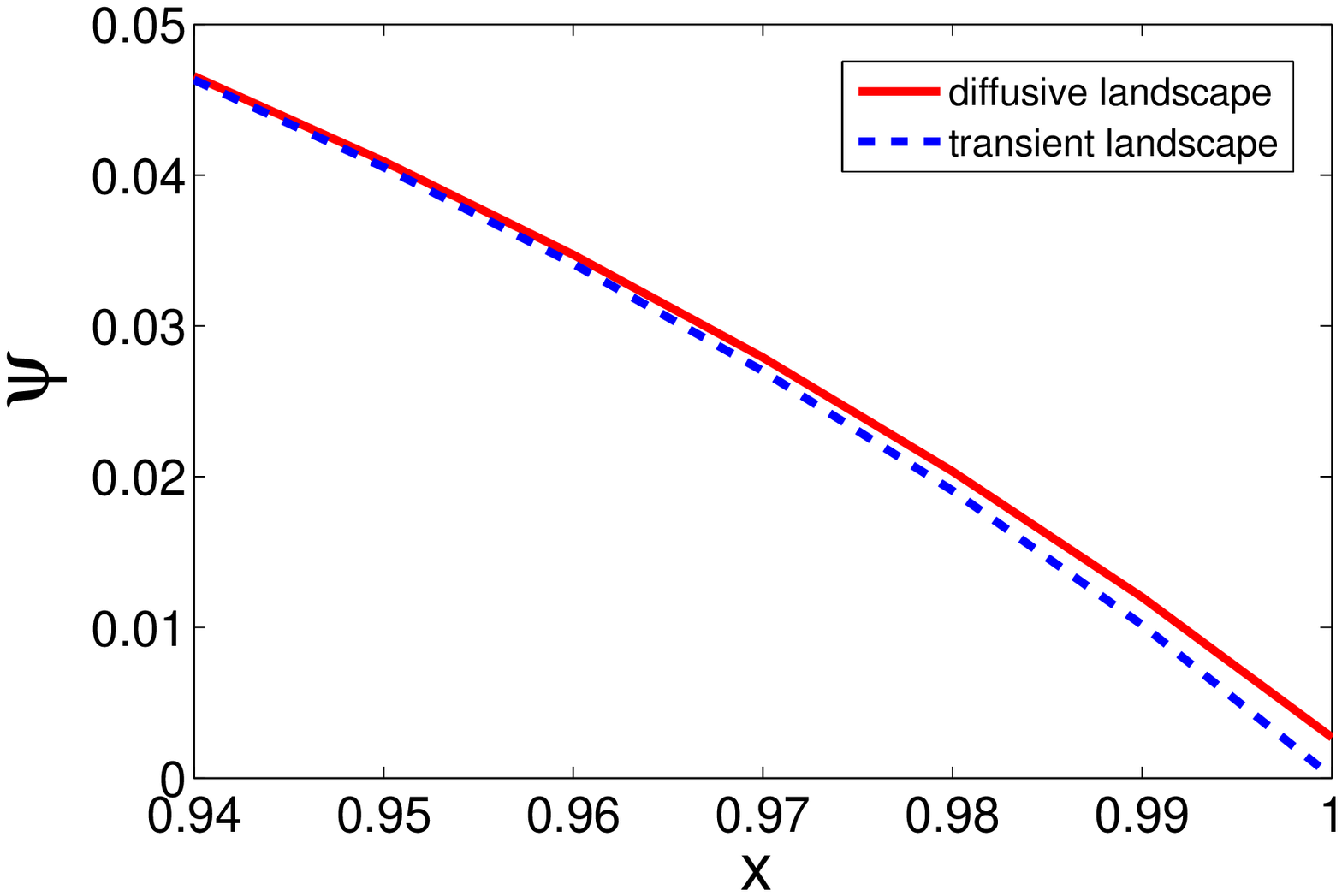}
\caption{}
\label{FIG_3}
\end{center}
\end{figure}

\begin{figure}
\begin{center}
\includegraphics[width=1\textwidth]{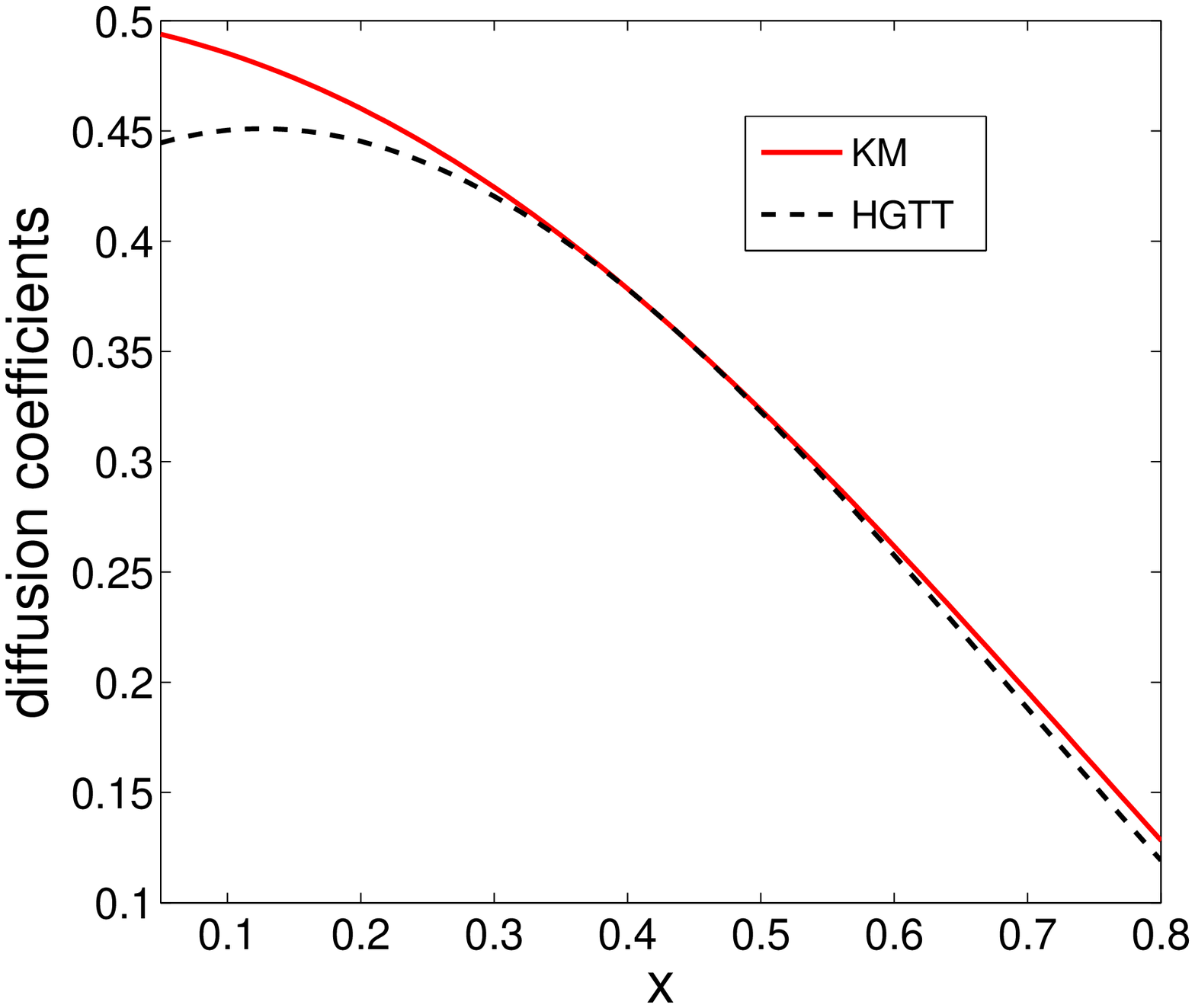}
\caption{}
\label{FIG_4}
\end{center}
\end{figure}

\end{document}